\documentclass{article}       
\usepackage[dvips]{graphicx}
\usepackage{amsmath}
\usepackage{epsfig}

\newlength{\doublefig} 

\numberwithin{equation}{section}

\begin{document}

\begin{center}
{\Large{Three-dimensional visualisation of tracks in OPERA nuclear emulsion films}}\\

\vspace*{2cm}
{J. Damet$^{a}$, E. Pereiro L\'opez$^{b}$, A. Sasov$^{c}$ }\\

 \vspace{5mm}
 {\small{
   $^{a}$ Laboratoire d'Annecy-le-vieux de Physique des Particules, \\
LAPP,9 Chemin de Bellevue - BP 110 F-74 941 Annecy-le-Vieux CEDEX - France \\
   $^{b}$ European Synchrotron Radiation Facility \\
	6, rue Jules Horowitz - BP 220 F-38 043 Grenoble - France\\
   $^{c}$ SkyScan, Vluchtenburgstraat 3, B-2630 Aartselaar, Belgium\\

 }}
\end{center}

\begin{abstract}
 The possibility of a three-dimensional visualisation/reconstruction of tracks in nuclear 
emulsion films using X-ray imaging is described in this paper. 
 The feasibility of the technique is established with experimental results.
\end{abstract}

\section{Introduction}
 Photographic emulsions have been proposed as detectors in 1952 for the study of high energy 
interaction in the cosmic radiation \cite{emulsion}. Nuclear emulsions are since used as high 
resolution tracking devices to study charged particles. The photographic method has been developed 
for the study of nuclear processes and used for discoveries concerning the mesons. 
  One of the early use of emulsion was in 1959 in the study of the composition of the 
Van Allen belts \cite{vanallen}.\\
 More recently, neutrino physics experiments used emulsions for unambiguous detection of charged 
tau lepton in $\nu_{\tau}$ interactions. Using nuclear emulsion detectors, the 
CHORUS experiment has set competitive limits on neutrino oscillations \cite{chorus} 
and the DONUT experiment has established the existence of the $\nu_{\tau}$ lepton \cite{donut}. 
The OPERA experiment \cite{opera} actually under construction at the Gran Sasso laboratory 
aims to confirm the neutrino flavour change.\\

 For our experimental study we used OPERA-like emulsions which are discribed in section \ref{sec:opera}. 
Results obtained with a new X-ray nano-tomograph system developped by SkyScan are discussed in section \ref{sec:skyscan}. 
Samples have been scanned at the ID19 beamline of the ESRF, results are shown in section \ref{sec:esrf}.\\

\section{OPERA nuclear emulsions}\label{sec:opera}
 
\begin{figure}[htb]
\begin{center}
\vspace*{-0.1cm}
\includegraphics[width=85mm]{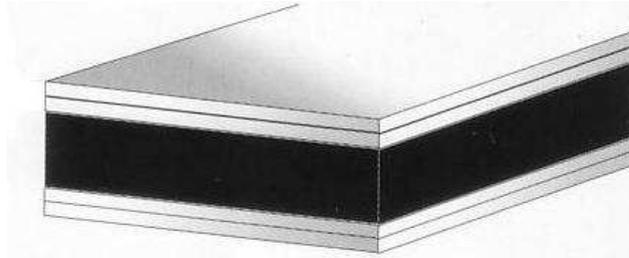}
\vspace*{-0.5cm}
\caption{Schematic diagram of an OPERA-like nuclear emulsion with an emulsion layer on both side of a plastic base.
A 1$\mu$m plastic layer is inserted in the middle of each emulsion layer.}
\label{fig:emulsion}
\end{center}
\end{figure}

 OPERA nuclear emulsion plates developped jointly by the FujiFilm company and the department of Fundamental 
Particle physics laboratory of the university of Nagoya are produced in an industrial mode. Dimensions of the 
sheets are $10 \times 12.5 \ {\rm{cm}}^2 $ large and 300 $\mu$m thin. Each plate is composed of a 50 $\mu$m 
layer of emulsion on both side of a 200$\mu$m thick plastic base. A schematic outline is shown in 
Fig.\ref{fig:emulsion}. Emulsion layers consist of gelatin containing silver bromide crystals.

 As a charged particle passes through the emulsion plate, silver bromide crystals are ionised and form 
after development metallic silver grains (smaller than a micrometer in diameter) opaque to light. 
Tracks hence appear as a serie of aligned silver grains.\\
 The read out of the nuclear emulsions in OPERA is done by automatited optical microscopes. 
The microspcope moves along the perpendicular axis of the emulsion. During the displacement, 
images are grabbed. The emulsion is thus virtually sliced into layers on which silver grains appear as black clusters and  
tracks inside an emulsion layer (micro tracks) are reconstructed by using a pattern recognition algorithm.\\
\par

 The X ray scanning technique can improve the reconstruction of microtracks with a precise localisation of 
the silver grains and can give a direct visualisation of the segments. Moreover this technique could lead 
to a precise counting of the grains along tracks and could thus be used as a powerful tool for particle 
identification. Indeed the location of grains with optical microscopes is limited by the focal depth of 
the objectives (~3$\mu$m in the case of the OPERA scanning system.)

\section{SkyScan-2011 NanoTomograph}\label{sec:skyscan}

\begin{figure}[htb]
\begin{center}
\vspace*{-0.1cm}
\includegraphics[width=108mm]{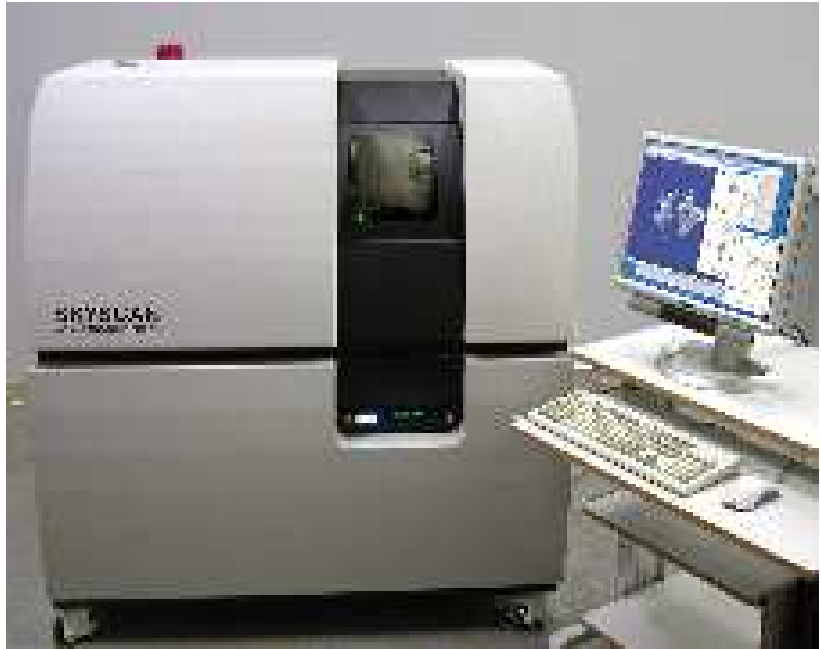}
\vspace*{-0.5cm}
\caption{The SkyScan-2011 system.}
\label{fig:skyscan2011}
\end{center}
\end{figure}

Precise tree-dimensional internal structural analysis of elements is now possible by X-ray nanotomography \cite{nanoscan}. 
 SkyScan-2011, illustrated in Fig.\ref{fig:skyscan2011}, is a compact laboratory X-Ray system 
for the nondestructive three-dimensional reconstruction of the objects internal microstructure 
with spatial resolution of 150 to 250 nanometers. No preparation, coating or vacuum treatment is needed. \\

\subsection{Results}

 We performed a test using an OPERA-like emulsion that had not been exposed to any beam. 
The emulsion had been checked using a simple optical microscope. Tracks from cosmics 
rays have been clearly identified although the emulsion has smaller dimension than 
standard OPERA sheets. The emulsion layers were no thicker than 30 $\mu$m.\\

\begin{figure}[h]
\begin{center}
\vspace*{-0.1cm}
\includegraphics[width=95mm]{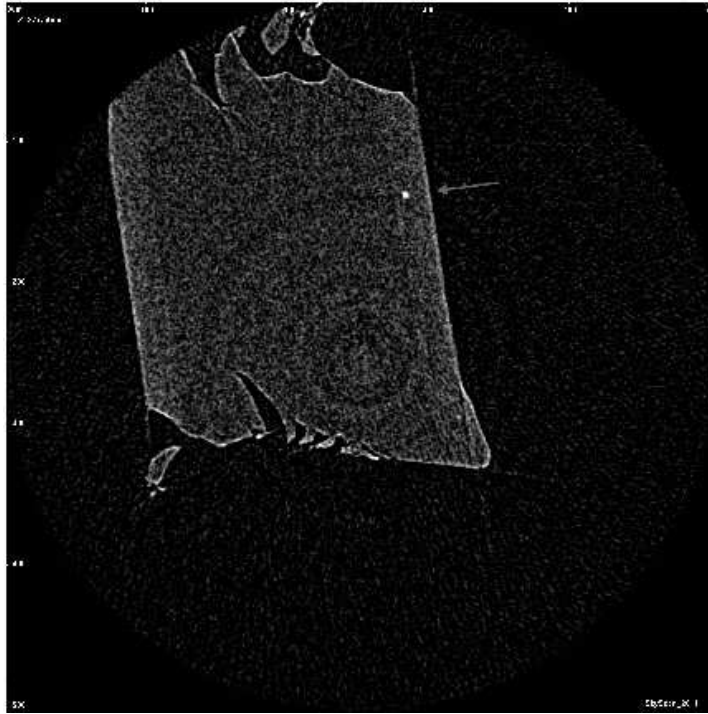}
\vspace*{-0.5cm}
\caption{Large cluster of silver grains in a layer of an OPERA-like emulsion film in a view of 500x500 $\mu$m$^2$.}
\label{fig:cluster}
\end{center}
\end{figure}

\begin{figure}[h]
\begin{center}
\vspace*{-0.1cm}
\includegraphics[width=95mm]{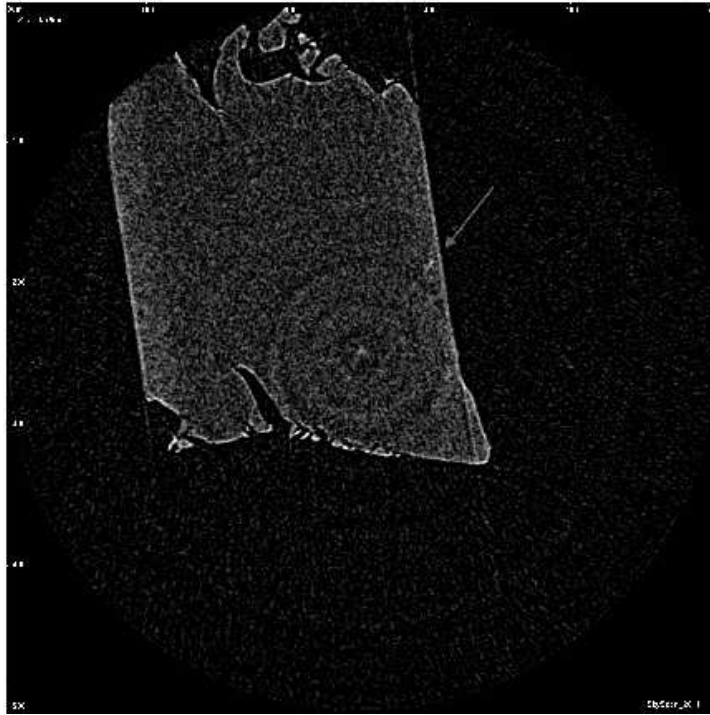}
\vspace*{-0.5cm}
\caption{Aligned sliver grains in a layer of an OPERA-like emulsion film in a view of 500x500 $\mu$m$^2$.}
\label{fig:line}
\end{center}
\end{figure}

\begin{figure}[h]
\begin{center}
\vspace*{-0.1cm}
\includegraphics[width=75mm]{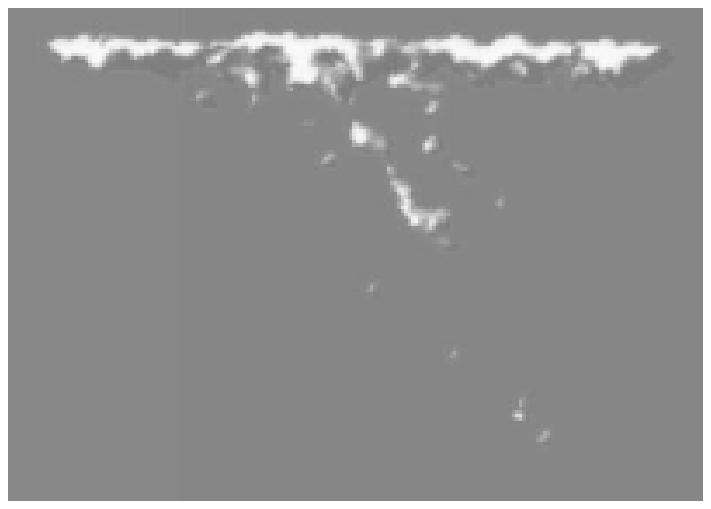}
\vspace*{-0.5cm}
\caption{Three-dimensional track from cosmics ray in the top side of an OPERA emulsion film.}
\label{fig:3D}
\end{center}
\end{figure}

 A small sample from the emulsion has then been scanned with the SkyScan-2011 nano-tomograph. 
Two typical kinds of shapes are identified in the emulsion: A cluster of silver grains with big 
absoption as shown in Fig.\ref{fig:cluster} and small aligned silver grains as 
shown in Fig.\ref{fig:line}. Both figures correspond to the same sample at different depth. 
Lighter thin bands (~30$\mu$m large) on the sides are emulsions layers, the darker large band in the 
middle (~190$\mu$m large) is the plastic base of the sheet. \\
 From the image data set, a three-dimensional reconstruction of the track inside the emulsion
layer is done as illustrated in  Fig.\ref{fig:3D}. The reconstructed microtrack is more likely 
to be a track generated by an alpha particle.

 The scanning time for a sample (1mm long) is one hour. The 3D reconstruction is done off-line.

\section{High-resolution Diffraction Topography Beamline}\label{sec:esrf}

 A second test has been performed at the high-resolution diffraction beamline 
ID 19 of the European Synchrotron Radiation Facility. A measurement with a spatial 
resolution of 0.7 $\mu$m has been performed.

\subsection{Results}
\begin{figure}[h]
\begin{center}
\vspace*{-0.1cm}
\includegraphics[width=90mm]{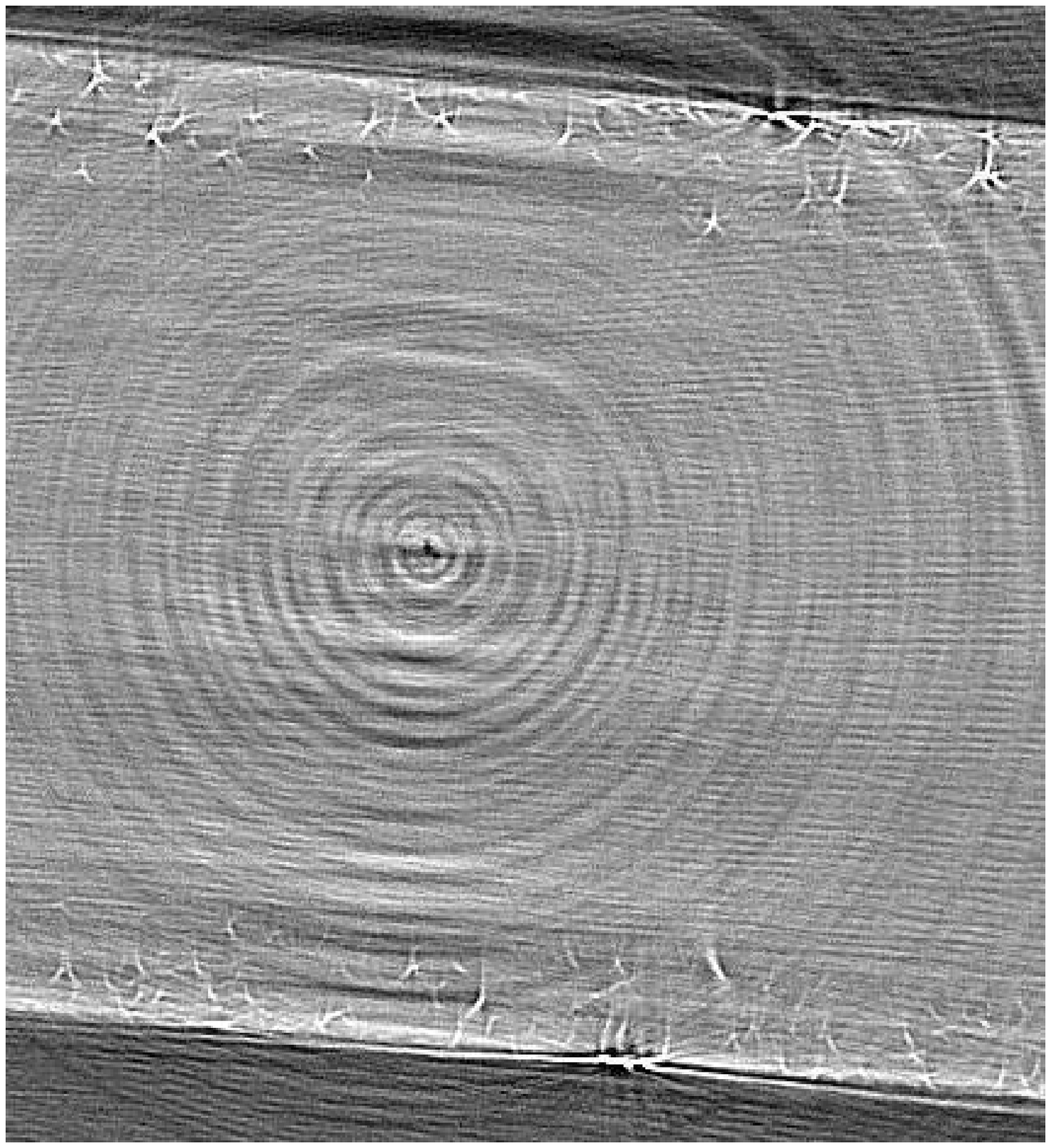}
\vspace*{-0.5cm}
\caption{Aligned sliver grains in both layers of an OPERA-like emulsion film.}
\label{fig:esrf}
\end{center}
\end{figure}

As shown on Fig:\ref{fig:esrf}, particle tracks can clearly be identified in both emulsion
layers, albiet the slight movement of the sample during the tomography
scan. They appear as brigth star shapes or lines due to this displacement. 
More precise measurements can be done by fixating the sample in the resin.\\

\begin{figure}[h]
\begin{center}
\vspace*{-0.1cm}
\includegraphics[width=90mm]{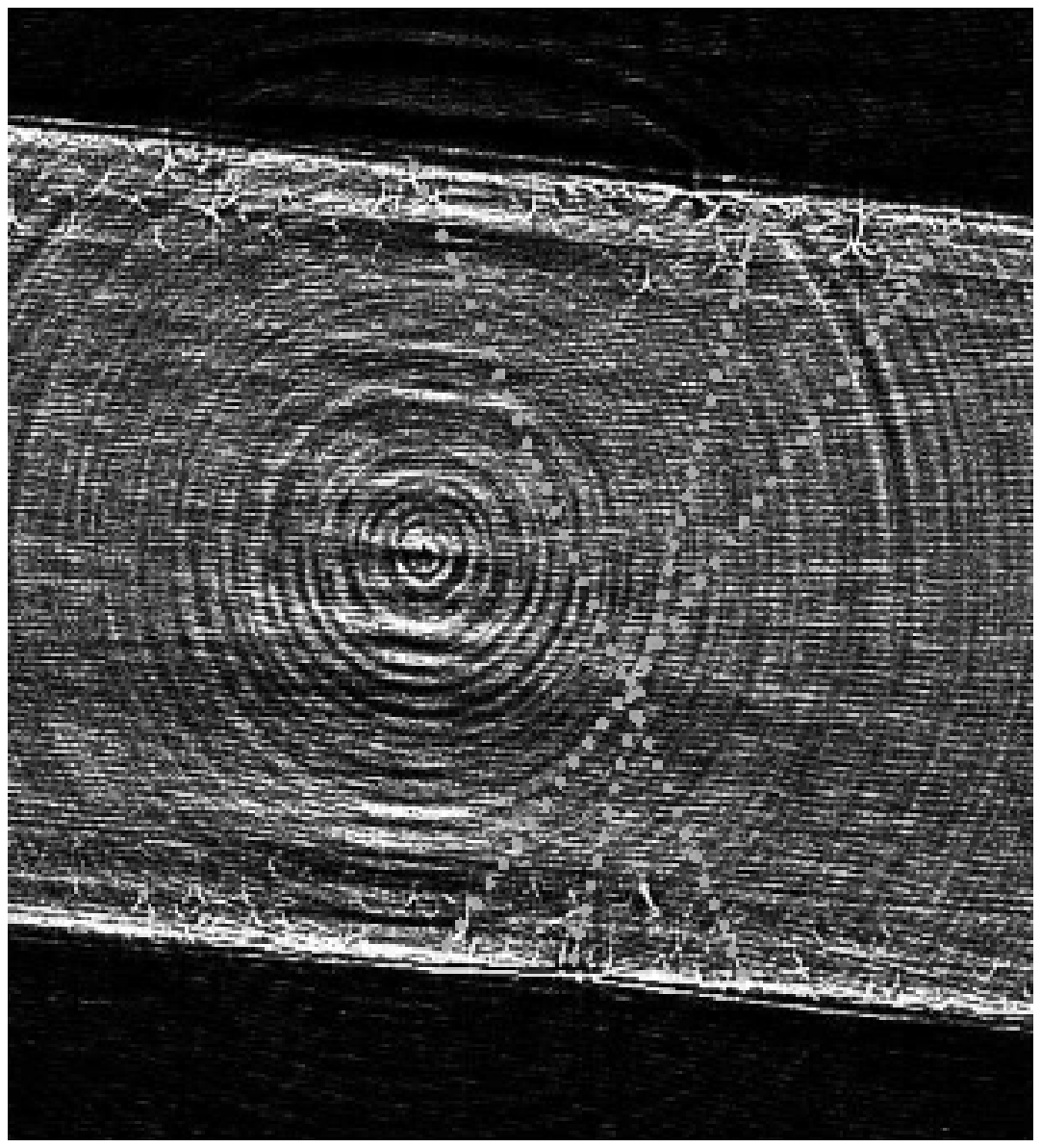}
\vspace*{-0.5cm}
\caption{Exemples of identified base-tracks, {\it{i.e.}} Connected segments of aligned sliver grains in both layers of an OPERA-like emulsion film.}
\label{fig:esrf-basetracks}
\end{center}
\end{figure}

 Once the image is treated base-tracks (connected segments from each emulsion sides) can be identified, 
as illustrated on Fig:\ref{fig:esrf-basetracks}\\
 Results of this test patently show track segments are identified in both layers of the emulsion. Few of these 
segements can be connected and are most likely tracks generated by cosmics.

\section{Summary and outlook}

 We demonstrated that tracks of charged particles inside nuclear emulsions
can be reconstructed in a three dimensional view using X-ray microtomography.
The present limitation of the size of samples to be scanned restricts the use of 
this technique, but results presented in this note indicate that this is an 
auspicious starting point as fast technological improvements are expected 
in the coming years.\\
 Modern X-ray imaging systems offer unprecedented possibilities of 3D reconstruction 
and measurements of charged tracks in nuclear emulsions.\\
 Performances to reconstruct tracks as required in OPERA would require intensive test 
beam compaign and could be described in future papers. 


\end{document}